# Dynamic mutation enhanced greedy strategy for wavefront shaping

Chuncheng Zhang[1], Xiubao Sui[1*], Zheyi Yao[1], Guohua Gu[1], Qian Chen[1], Zhihua Xie[2], Zhihua Xiong[2], Guodong Liu[2]

1. School of Electronic Engineering and Optoelectronic Technology, Nanjing University of Science and Technology, Nanjing 210094, China
2. Key Lab of Optic-Electronic and Communication, Jiangxi Science and Technology Normal University, Nanchang 330013, China

*Corresponding email: sxb@njust.edu.cn

**Abstract:** Optical focusing through scattering media has important implications for optical applications in medicine, communications, and detection. In recent years, many wavefront shaping methods have been successfully applied to the field, among which the population optimization algorithm has achieved remarkable results. However, the current population optimization algorithm has some drawbacks: 1. the offspring do not fully inherit the good genes from the parent. 2. more efforts are needed to tune the parameters. In this paper, we propose the mutate greedy algorithm. It combines greedy strategies and real-time feedback of mutation rates to generate offspring. In wavefront shaping, people can realize high enhancement and fast convergence without a parameter-tuning process.

## 1. Introduction

Electromagnetic waves have scattering properties as they propagate[1, 2]. For example, the signal of WIFI is essential in modern life. It is excellent when the router is in the same room as the networked devices[3]. But it is poor when they are in different rooms because electromagnetic waves are scattered when they penetrate walls. The signal is rapidly attenuated and does not reach the receiving end[4]. Light, an electromagnetic wave, scatters strongly when passing through scattering media with an uneven refractive index, such as smoke, biological tissue, and multimode fibres[5]. The intended light information is transmitted incorrectly. These phenomena seriously hinder the development of detection imaging, biomedicine, communications, and other fields[6].

To overcome these problems, Vellekoop first proposed continuous sequence algorithms (CSA) to modulate the wavefront phase information of light to achieve optical focusing through a scattering media in 2007 [7]. Since then, wavefront shaping methods have attracted a lot of interest from researchers. Over the last few decades, many excellent iterative algorithms have been proposed to improve the performance of wavefront shaping, such as the differential evolutionary algorithm (DEA)[8], genetic algorithms (GA)[9], and simulated annealing algorithms (SAA)[10], etc. For all these classical algorithms, there are some problems. CSA is simple and efficient but susceptible to noise. As global optimization algorithms, GA and DEA are not affected by noise but suffer from more parameters for modulation and locally optimal solutions. SAA accepts poor solutions with some probability of jumping out of the local optimum, but its convergence becomes slow.

The greedy algorithm always chooses the best solution when solving the problem [11-15]. Although it has a reasonable convergence rate, the choice is only the local optimal solution. This solution is not considered from the iteration of the population and is not globally optimal. In many other problems, greedy algorithms, such as the traveling salesperson problem, may even produce the worst solution. The nearest neighbor is computed for each city, and this approach may produce the worst travel distance[16]. Population optimization algorithms use the exchange and cooperation of information between populations due to simple and limited individual interactions to achieve optimization[17-19]. Then they can get outstanding robustness, stability, and adaptability, e.g., GA, DEA, and PSO. In general, they avoid the above problems. However, the calculation of population iterations is relatively complex, requiring more parameters such as the GA containing population size, initial variation rate, final variation rate, and decay factor. At the same time, their rate of convergence can be affected[20]. Generating the next generation, such as crossover and difference, is somewhat stochastic. Offspring are not always beneficial to the population [18].

In this paper, a new population optimization algorithm, the mutate greedy algorithm (MGA), is constructed by combining a greedy algorithm with a mutation operation. The MGA selects the best population as the next generation's parent and mutates it to obtain a new population. In wavefront shaping, the mutation rate is generally a fixed value or a decaying value that varies with the number of iterations[21]. Generating new populations may significantly reduce the efficiency and diversity of the population's optimization. The final result may fall into a locally optimal solution. Hence we need a reasonable variation rate[22]. A dynamic variation operation is utilized to balance the contradiction between the greedy strategy and the diversity of the population: the mutation proportion of the mask of the optimal population is dynamically adjusted, i.e., the adaptive mutation, to expand the population diversity in each iteration. In theory, MGA has two advantages :1. It has a faster convergence speed than GA, SAA, and CSA, etc. 2. It only needs to adjust the number of populations, inheriting the simplicity and efficiency of the greedy algorithm. The experimental and simulation results verify the correctness of the theory. Also, we find that the population size has less influence on the MGA. This approach will further advance the application of wavefront shaping in image transmission and light focusing.

## 2. Principle

*2.1 Wavefront shaping*

The transmission of incident light through a scattering medium is deterministic. Mathematically, the scattering medium can be represented by the transmission matrix $t_{mn}$, which $m, n = 1, ..., N$. $t_{mn}$ The light intensity of desired focus $I_m^{out}$[7]:

$$I_m^{out} = \left|\sum_n^N t_{mn} A_n e^{i\varphi_n}\right|^2 \tag{1}$$

$A_n$ and $\varphi_n$ are the amplitude and phase of the n-th controlled segment of the incident light. The focusing algorithm aims to find the best $\varphi_n$. The performance of focusing is defined as follows[23]:

$$\eta = \frac{I_m^{out}}{\langle I_0 \rangle} \tag{2}$$

where $\langle I_0 \rangle$ is the average intensity of the outgoing light, and $\eta$ is the enhancement of the focusing point. The theoretical enhancement $\eta_{ideal}$ is determined by the modulable segment N of the optical field[7]:

$$\eta_{ideal} = \pi(N-1)/4 \tag{3}$$

*2.2 Variational greedy algorithm*

The specific flow of the greedy variant algorithm is shown in Figure 1. Each population is a mask on a liquid crystal phase modulator (SLM) in the mutated greedy algorithm. The algorithm ranks the fitness of the population and finds the best population.

$$G_{best} = \text{Max}(G_i) \quad i = 1,2,3 ... M \tag{4}$$

$G_i$ is the individual population, and $M$ is the maximum number of populations. Calculate the $G_{best}$ which corresponds to the light field $I_{best}$.

The Pearson coefficient (PCC) is a classical image quality evaluation metric. It is often used as a criterion for similarity judgment and as a cost function in optimization[24]:

$$\gamma = \frac{\sum_{i=1}^{U} \sum_{j=1}^{V} [X(i,j)-\bar{X}][Y(i,j)-\bar{Y}]}{\sqrt{\sum_{i=1}^{U} \sum_{j=1}^{V} [X(i,j)-\bar{X}]^2} \sqrt{\sum_{i=1}^{U} \sum_{j=1}^{V} [Y(i,j)-\bar{Y}]^2}} \tag{5}$$

$i$ and $j$ denote the coordinates of the image, the $X$ and $Y$ denote the target image and the comparison image, respectively, and $\bar{X}$ and $\bar{Y}$ are the value of $X$ and $Y$.

PCC calculates the correlation between the reconstructed and preset light fields. PCC ranges from (-1, 1) and is taken as an absolute value to meet the practical needs of the algorithm. The value of PCC is a simple and efficient way to determine the difference between the focal point and the ideal focal point. It can guide the algorithm to update the appropriate rate of mutation $R$. The population's diversity will increase to prevent the algorithm from falling into a local optimum solution. The specific update method of $R$: a logarithmic function is used to optimize the value of PCC and construct an adaptive rate of

mutation[21]:

$$R = \begin{cases} -\log(10^{-4})/C, |\alpha| < 10^{-4} \\ -\log(|\alpha|)/C, \text{else} \end{cases} \quad (6)$$

Based on $R$, $G_{best}$ is randomly mutated N times. A new population is constructed. The above operation is repeated, and the MGA finds the current optimal solution and performs a mutation up to the maximum number of iterations.

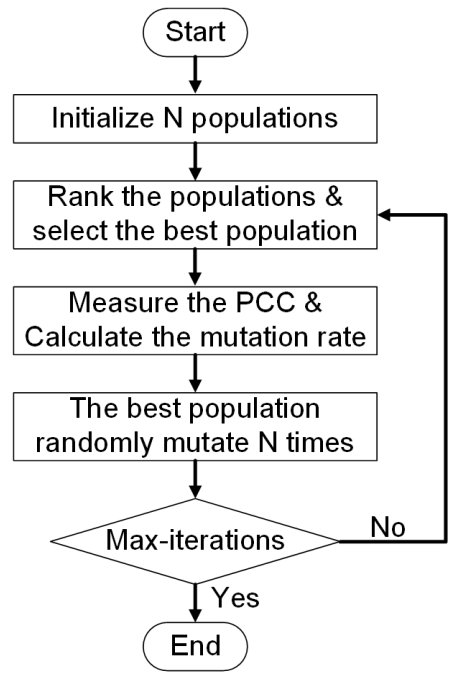

Figure 1. The flow chart of MAG.

## 3 Experimental results and discussion

### 3.1 Parameters and system of the experiment

We compared MGA with other classical wavefront algorithms to demonstrate the advantages of convergence speed and without parameters to adjust. These algorithms are GA, CSA, and SAA, and their parameters are shown in Table 1. MGA only needs to choose the number of populations, which impacts the convergence speed and the final enhancement. Its value is 20 to balance the conflict between the diversity of the population and the optimal solution. The constant C is taken as 40. For the GA, the population size is 40, initial and final mutation rates are 0.1 and 0.001, respectively, and the decay factor of mutation is 200. SAA's initial and final temperatures are 99 and 0.01, respectively. The length of its Markov chain is 20. For CSA, the step size of the phase modulation is π/10. The experimental results of each simulation were averaged over 100 independent optimizations. For all algorithms, the phase mask of size N is set to 256, i.e., a phase mask of size 16×16. The phase takes a range of values from 0-2π.

Table 1 Experimental parameters

| Algorithm | Initial Parameters |
| --- | --- |
| MGA | 1. Population = 20 <br> 2. C = 40 |
| GA | 1. Population = 40 <br> 2. Initial mutation rate = 0.1 <br> 3. final mutation rate = 0.001 <br> 4. Decay factor = 200 |
| SAA | 1. Initial temperature = 99 <br> 2. Final temperature = 0.001 <br> 3. Length of Markov chain = 20 |
| CSA | 1. Size of step = 10 |

The experimental setup of the system is shown in Figure 2. The light source is a continuous 532-nm laser (LCX-532S, Oxxius, 80 mW). The BE (GBE20-A-20 ×, Thorlabs, USA) expands the light. The beam's polarization is adjusted by Polarizers P. The SLM modulates and reflects the light (PLUTO-2-vis-096, HOLOEYE; pixel pitch: 8 µm) and enters a 4F telescope system. The 4F telescope system is used to adjust the magnification of the target. Finally, a CCD (BFS-U3-04S2M-CS, Point Gray; pixel pitch: 6.9 µm) receives the beam after the light has been scattered by the scatterer S. There are 16×16 controlled segments involved in the illuminated area of the SLM which is a pure phase- type. A segment is a group of 20×20 adjacent SLM pixels. The central pixel of the CCD is chosen as the feedback position, and its surrounding 300×300 pixels are detected during the process of experimental optimization.

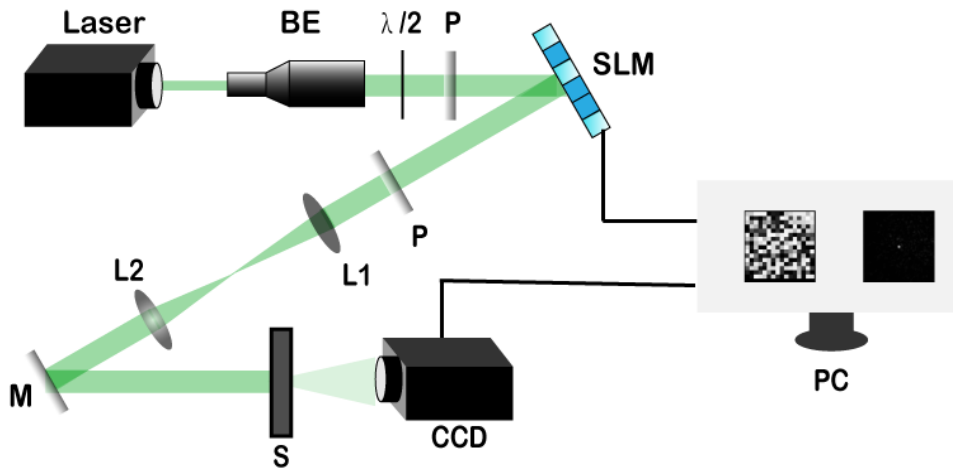

Fig. 2 The optical system for experiments. BE: beam expander; L1, L2: lens, $\lambda/2$: half-wave plate, P: polarizer, M: mirror; PC: a personal computer, SLM: spatial light modulator.

*3.2 Simulation.*

In computational optical imaging, noise is inevitable, including read-in noise from the camera, interference from ambient light, and the effects of the self-referencing optical path. The without, low, medium, and high noise are represented as $0<I0>$, $0.3<I0>$, $0.6<I0>$, and $0.8<I0>$. They can affect the performance of the focus. Figure 3 clearly shows the focus of each algorithm in different noise. $<I0>$ is the average intensity of the initial background.

Figure 3 clearly shows the performance of all algorithms with different levels of noise. As shown in Figure 3(a), the MGA, GA, and CSA all reach nearly the values of their theoretical enhancement at 20,000 measurements without noise. Due to its Monte-Carlo strategy, the SAA is around 80% of the theoretical value, but it still has room to rise. At 2000 measurements, the MGA exceeds 50% of the theoretical value. The GA and SAA only reach 30% of the theoretical value. The MGA's speed of convergence is 2/3 faster than theirs. The enhancement of CSA is less than 20% of the theoretical value of less than 20%, and the enhancement of MGA is two times stronger than it. As the increases of noise, not only the convergence rate of CSA gradually decreases, but its value of final enhancement is also greatly affected. Especially in the early stage, CSA barely converges. Although CSA has only one parameter to adjust, it is hardly applicable in practical applications. MGA and GA as global optimization algorithms. SAA randomly accepts poor solutions. Their final enhancements are almost unaffected by noise, as shown in Figures 3 (b), (c), and (d). The convergence of MGA is still the fastest among all algorithms with noise. At 2000 measurements, the convergence of MGA, GA, and SAA is as fast as before. Meanwhile, the MGA achieves an enhancement of about 90% theoretical value at 8000 measurements. But GA can only reach about 70% of the theoretical value, and SAA is less than 60%. From the simulation results, the MGA seems more suitable for practical applications than the GA, SAA, and CSA.

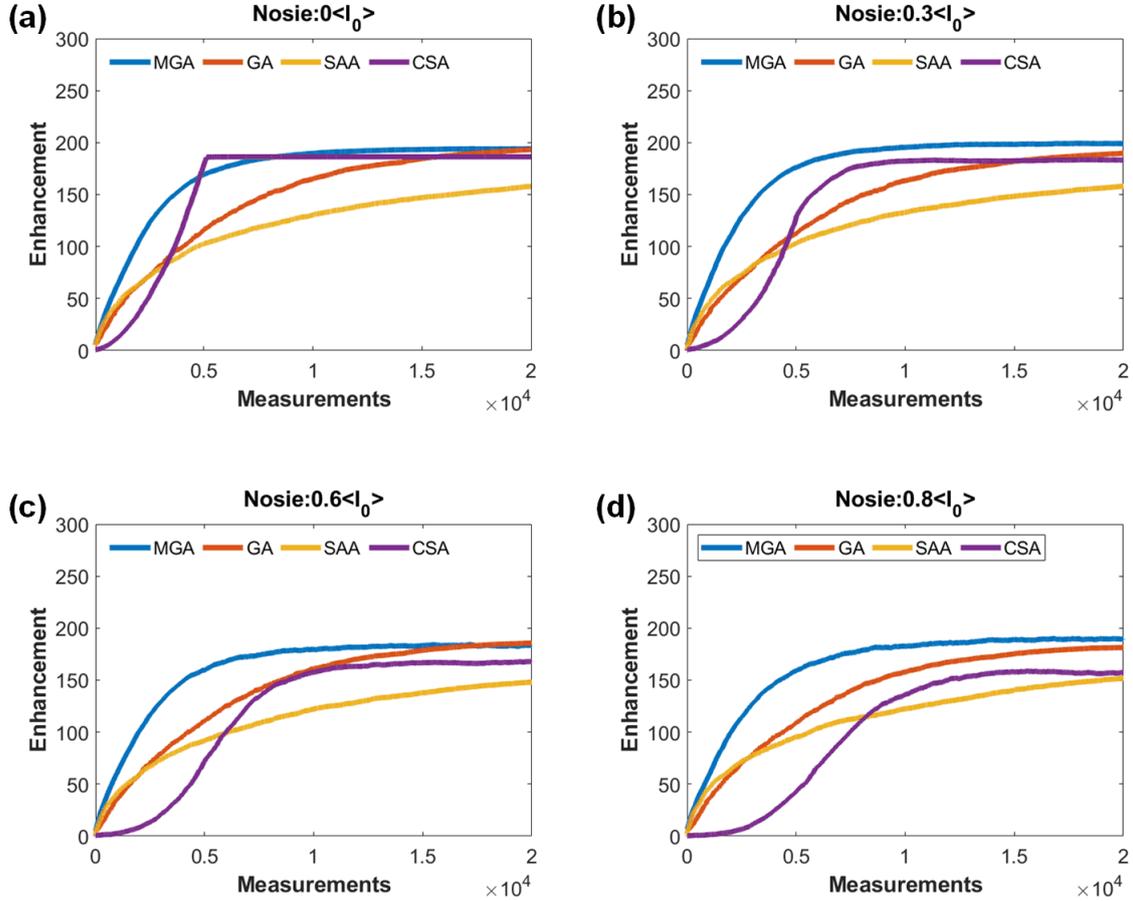

Fig. 3. Simulation results on the algorithm with different noise levels. (a), (b), (c), and (d) represent $0 < I_0 >$, $0.3 < I_0 >$, $0.6 < I_0 >$, and $0.8 < I_0 >$, respectively. and $0.8 < I_0 >$ respectively.

The parameters that MGA can adjust are the number of populations. If it has more populations, the diversity of the population is better. However, the more populations it has, the more measurements it needs at every iteration. Thus the convergence rate of MGA will be slow. To find the optimal population, experiments were conducted with population sizes of 10, 20, 40, 60, and 100, as shown in Figure 4. The smaller the number of populations, the faster the convergence of the algorithm, but small populations are more likely to fall into local optimal solutions, resulting in a low final enhancement. The curve with a population of 10 converged fastest in the early part, reaching saturation first. However, due to the small size of the population, it lacks diversity. Its final enhancement is relatively not the highest of all of them. The curves with populations of 20, 40, 60, and 100 follow the above theory, and their convergence rate gradually decreases. It is worth noting that a more significant number of populations, such as 60 and 100, hinders the efficiency of the greedy strategy. Even the final enhancement for a population of 100 is not as high as that of a population of 10. The final enhancements of the population with 20 are similar to the population with 40, but the convergence rate of the population with 20 is faster than 40. Taking the above into account, the population size of 20 is suitable for the paper.

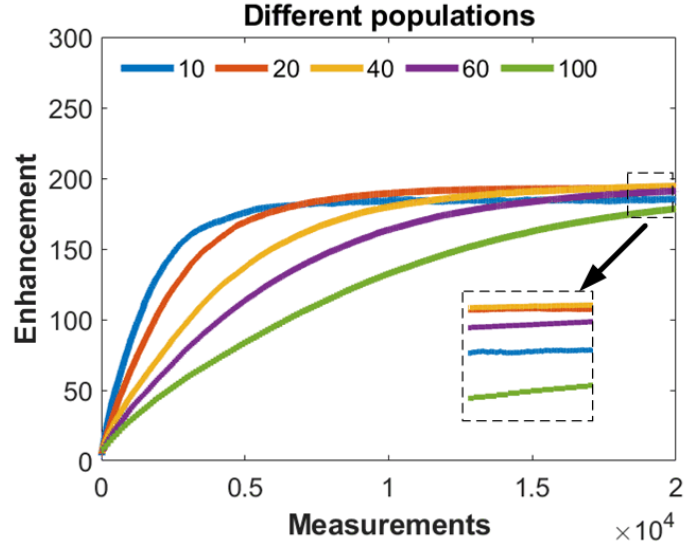

Fig. 4 Enhancements of MGA with population sizes of 10, 20, 40, 60, and 100, respectively.

*3.3 Experiments*

In this section, we show the experimental results and analyze them. The parameters of all algorithms are consistent with the simulation. All algorithms were performed for 20,000 measurements, and their curves of enhancements and final focus are shown in Fig. 5. The values of enhancements for 8000 measurements and 20,000 measurements are shown in Table 2. As shown in 5(a), the enhancements of MGA, GA, SSA, and CSA are 50.29, 44.18, 10.53, and 29.36, respectively, at the early optimization stage (8000 measurements). The growth trend of the enhancements of all algorithms is mainly similar to the simulated results. However, the performance of the individual optimization algorithms (CSA and SAA) is significantly affected by experimental errors, such as system noise, excess modulation amplitude of the self-referencing system, and deviation of the optical path. Their enhancements are much lower than those of MGA and GA. The GA has a high value of enhancement because multiple individuals are optimized simultaneously, enhancing the diversity of the algorithm. However, the crossover strategy in GA is flawed. The two individuals of the parent generation cross randomly, and not every mask of the next generation can inherit the parent's good genes. Some of the new masks may contribute negatively to the enhancement. MGA has obtained the highest value of enhancement than other algorithms. It inherits the feature that all offspring are optimal parents in individual optimization while combining the mechanism of multiple populations in population optimization to resist random experimental errors. As a result, MGA has both high enhancement and the ability to converge quickly in the early stage. At about 14,000 measurements, the enhancement of MGA is 58.66. It has converged to the maximum value of enhancement in the experimental setting. At the time, SAA has long fallen into a local optimum due to individual evolution. The CSA is affected by the experimental noise, and the value of the enhancement decreases instead of rising. GA may have better population diversity than MGA because of the crossover strategy and can still grow later. However, GA's crossover strategy may inherit poor genes, and the rise of enhancement is almost negligible during the 14,000-20,000 measurements. At 20,000 measurements, the enhancements of MGA, GA, SSA, and CSA are 58.66, 48.68, 10.53, and 31.45, respectively, and the final focus points are shown in 5(b-e). The values of the final enhancement of MGA are 20.5%, 457.1%, and 86.52% higher than those of GA, SAA, and CSA, respectively. The experimental results demonstrate the superiority of MGA in the field of wavefront shaping.

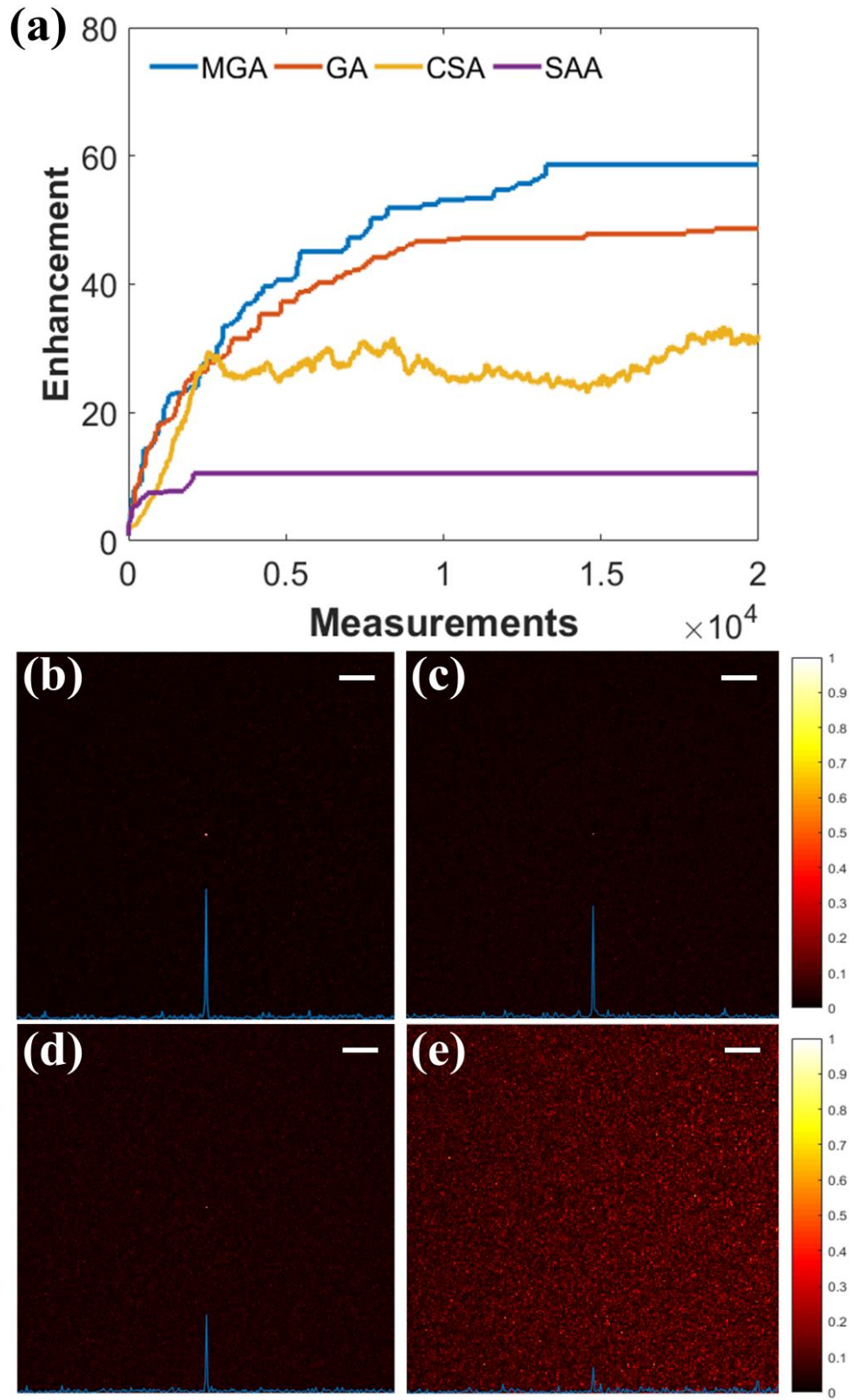

Fig. 5 Experimental results after 20000 measurements. The scale bar is 200 μm, denoting a length of 28.99 CMOS pixels. (a) Experimental enhancement curves of the MGA, GA, CSA, and SAA. (b-e) Results of MGA, GA, CSA, and SAA, respectively.

Table2 Experimental results of MGA, GA, SAA, and CSA.

| Number of measurements | MGA | GA | SAA | CSA |
|---|---|---|---|---|
| 8000 | 50.29 | 44.18 | 10.53 | 29.36 |
| 20000 | 58.66 | 48.68 | 10.53 | 31.45 |

The variable parameter in MGA is only its population size, and we conducted experiments for different population sizes to verify the performance of the algorithm. As shown in Figure 6, we set the number of populations to 10, 20, 40, 60, and 100, and the values of their enhancement were 52.16, 58.66, 56.10, 53.42, and 52.16, respectively. The experimental results are in general agreement with the simulation. From the experimental results, the smaller the number of populations, the faster the convergence speed of the algorithm in the early stage. However, the smaller the number of populations, the lower the diversity of its. The enhancement of MGA is easy to fall into the local optimum, as shown in Table 3. After 20,000 measurements, the number of measurements for which the MGA reached its current maximum was 11,460, 13,020, 15,180, and 15,680 for population sizes of 10, 20, 40, and 60, respectively. The enhancement of MGA tends to increase at 20,000 measurements when the number of populations is 100, but its value is the smallest than others. From the experimental results, MGA can balance the contradiction between greedy strategy and population diversity when the population size is 20.

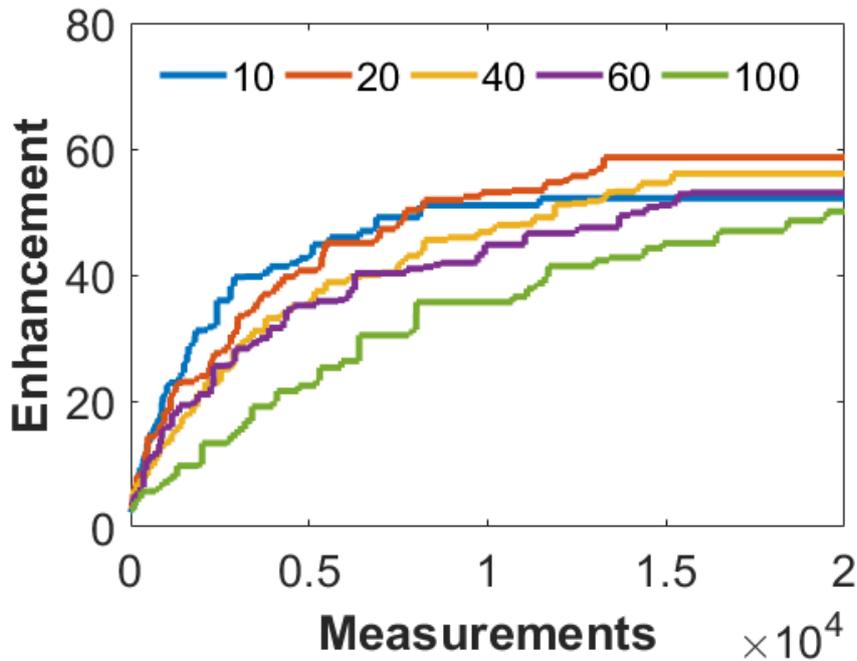

Fig.6 Experimental Enhancements of MGA with population sizes of 10, 20, 40, 60, and 100, respectively.

Table 3 The final enhancement in the number of different populations and the number of measurements of the first arrival

| Population size | 10 | 20 | 40 | 60 | 100 |
|---|---|---|---|---|---|
| Number of measurements of the first arrival | 11460 | 13020 | 15180 | 15680 | 20000 |
| Final enhancement | 52.16 | 58.66 | 56.10 | 53.42 | 52.16 |

## 4. Conclusion

In summary, we have demonstrated that MGA can converge quickly and achieve outstanding focusing results. Firstly, the algorithm's performance is unaffected by a high level of noise. Then, MGA can focus quickly due to its excellent capability of convergence. Despite these advantages, the performance of MGA is open to debate when real-time imaging is needed. To address this issue, a digital microlens device (DMD) will be considered, which has refresh rates of 100 kHz in our perception. It has the potential to solve the problem in whole or in part. In practical applications, the reflection mode is more suitable for various scenarios than the transmission mode. Our next workshop will be devoted to combining DMD and reflection mode. It is hoped that the methods and experimental results in this paper are expected to promote further development in wavefront shaping.

**Compliance with ethics guidelines**

All authors declare that they have no conflict of interest or financial conflicts to disclose.

**Data availability**

Data underlying the results presented in this paper are not publicly available but may be obtained from the authors upon reasonable request.

**Acknowledgments**

This work was supported by the National Natural Science Foundation of China ((No. 62105152), Key Research & Development programs in Jiangsu China(Grant no. BE2018126), Fundamental Research Funds for the Central Universities (Grant NO. JSGP202202, 30919011401，30920010001), Leading Technology of Jiangsu Basic Research Plan (BK20192003), The Postgraduate Research & Practice Innovation Program of Jiangsu Province(KYCX22_0411), The Open Foundation of Key Lab of Optic-Electronic and Communication of Jiangxi Province (NO.20212OEC002).